\documentclass[prl,twocolumn,a4paper,amsmath,floatfix]{revtex4}
\usepackage[utf8]{inputenc}
\usepackage{graphicx}
\usepackage{amsmath}
\usepackage{relsize}
\usepackage{amssymb}

\usepackage{color}
\definecolor{strawberry}{rgb}{1.0,0.0,0.5}

\definecolor{royallBlue}{rgb}{0.003921569,0.070588235,0.474509804}

\begin{document}

\title{Crystal polymorph selection mechanism of hard spheres hidden in the fluid \\}

\author{Gabriele M. Coli$^1$,  Robin van Damme$^1$, C. Patrick Royall$^{2,3,4}$, Marjolein Dijkstra$^1$}
\affiliation{$^1$Soft Condensed Matter, Debye Institute of Nanomaterials Science, Utrecht University, Princetonplein 1, 3584 CC Utrecht, Netherlands \\
$^2$Gulliver UMR CNRS 7083, ESPCI Paris, Universit\'{e} PSL, 75005 Paris, France \\
$^3$H.H. Wills Physics Laboratory, Tyndall Avenue, Bristol, BS8 1TL, UK \\
$^4$ School of Chemistry, University of Bristol, Cantock's Close, Bristol, BS8 1TS, UK \\}


\maketitle

\textbf{Nucleation plays a critical role in the birth of crystals and is associated with a vast array of phenomena such as protein crystallization and ice formation in clouds.
Despite numerous experimental and theoretical studies, many aspects of the nucleation process like the polymorph selection mechanism in the early stages are far from being understood. Here, we show that the excess of particles in a face-centred-cubic (fcc)-like environment with respect to those in a hexagonal-close-packed (hcp)-like environment in a crystal nucleus of hard spheres as observed in simulations and experiments \cite{pusey1989structure,palberg1997colloidal,cheng2001colloidal,filion2011simulation,russo2012microscopic,sandomirski2011} can be explained by the higher order structure in the fluid phase. We show using both simulations and experiments that, in the metastable fluid phase, fivefold symmetry clusters -- pentagonal bipyramids (PBs) -- known to be inhibitors of crystal nucleation \cite{frank1952supercooling,taffs2016role}, transform  into a different cluster -- Siamese dodecahedra (SDs). Due to their geometry, these clusters form a bridge between the fivefold symmetric fluid and the fcc crystal, thus lowering its interfacial free energy with respect to the hcp crystal, and shedding new light on the polymorph selection mechanism.} \newline

\begin{figure*}
\centering
\includegraphics[width=1.0\textwidth]{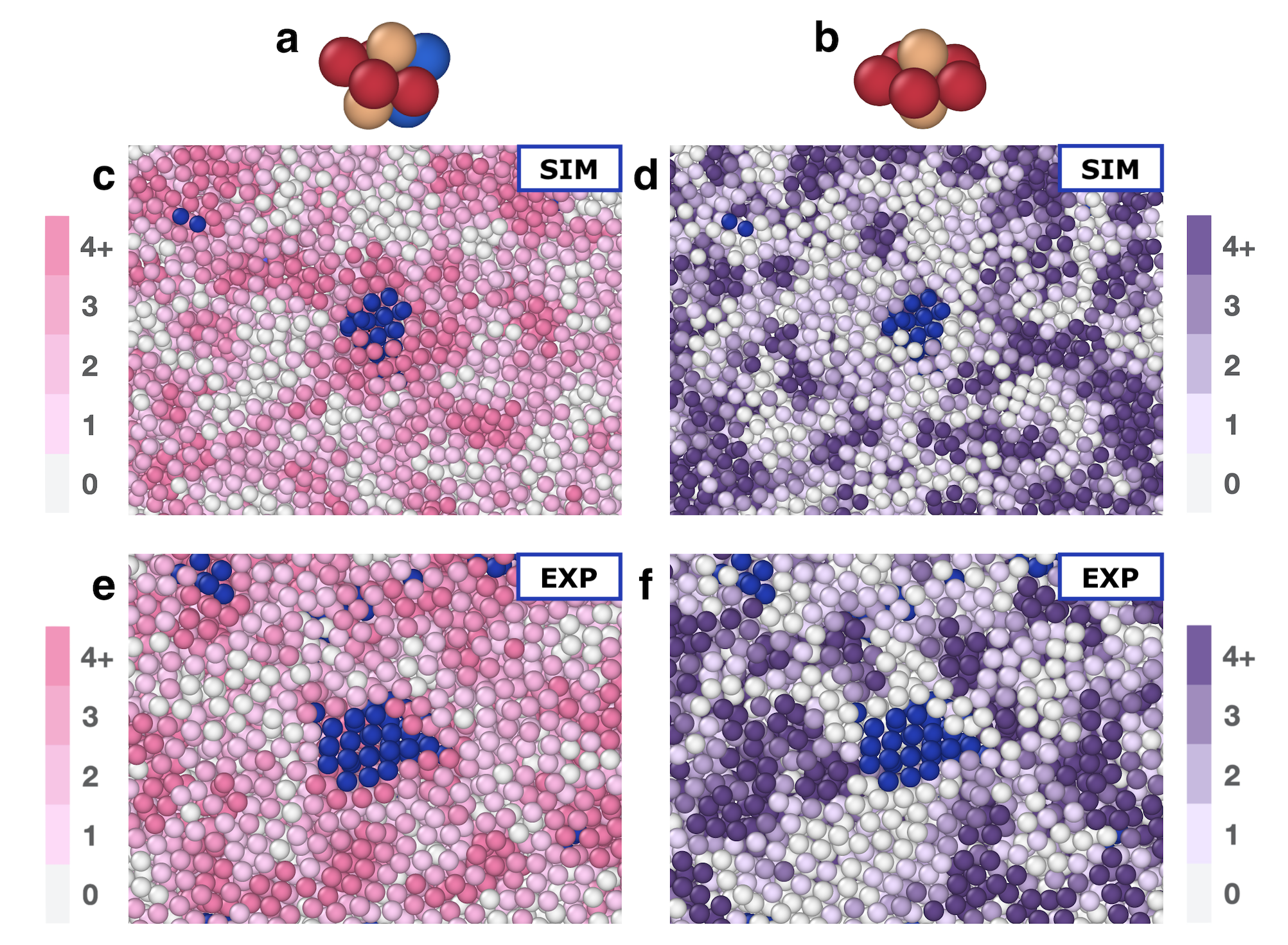}
\caption{\textbf{Typical configuration of a crystal nucleation event of hard spheres} 
(a-b) Rendering of the arrangement of particles in (a) a Siamese Dodecahedron (SD) and (b) Pentagonal Bipyramid (PB) cluster. The colour coding is explained in the text. 
(c-d) Cut-through image of an early-stage  nucleation event generated by MD simulations. Crystal-like particles are coloured blue, while fluid-like particles are coloured following the scale bar on the left (c) or right (d) depending on the number of SD (c) or PB (d) clusters each particle belongs to. The system is simulated at constant pressure $\beta P \sigma^3 = 13.80$, \emph{i.e.} starting from a fluid configuration at average effective packing fraction $\langle \phi_\mathrm{eff} \rangle = \frac{\pi}{6} \langle \rho \rangle \sigma_\mathrm{eff}^3 = 0.541$, where $\langle \rho \rangle$ is the average number density of the system, while $\sigma_\mathrm{eff}$ is the effective diameter of the particles (see Methods for the calculation of the effective packing fraction). (e-f) Experimental configuration at effective packing fraction $\phi_\mathrm{eff} = \frac{\pi}{6} \rho \sigma_\mathrm{eff}^3 = 0.541$.}
\label{fig:fig2}
\end{figure*}

Understanding nucleation is important in many research fields such as determining the molecular structure of proteins through crystallization, drug design in the pharmaceutical industry, ice crystal formation in clouds -- the largest unknown in the earth's radiative balance and thus crucial in the context of climate change and weather forecasts -- and crystallization of colloidal and nanoparticle suspensions with application perspectives in catalysis, opto-electronics, and plasmonics \cite{sear2007nucleation,palberg2014crystallization}. 

However, nucleation is extremely challenging to study in molecular systems as it is a stochastic and rare process,  and the sizes of the crystal nuclei are often rather small and the nuclei grow out extremely fast when they exceed their  critical size. An additional obstacle is 
that, for most substances,  different crystal polymorphs may compete during nucleation. This phenomenon is of key importance in pharmaceutical sciences and applications as the crystallization of the ``undesired'' 
polymorph may for instance lead to  neurodegenerative disorders such as Alzheimer's disease or eye cataract, or to reduced solubility/efficacy and even toxicity of certain drug compounds \cite{ohm1995apolipoprotein,bauer2001ritonavir}. 

Recently, impressive strides have been made in the experimental observation of early-stage crystal nucleation by using atomic-resolution in situ electron microscopy, showing the observation of different nucleation pathways to different crystal polymorphs of proteins \cite{van2018molecular}, pre-nucleation clusters in metal organic frameworks \cite{xing2019atomistic}, early-stage nucleation pathways of FePt nanocrystals that go beyond classical nucleation theory and  non-classical scenarios \cite{zhou2019observing}, amorphous precursors in protein crystallization \cite{houben2020mechanism}, featureless and semi-ordered clusters of NaCl nanocrystals \cite{nakamuro2021capturing}, and reversible disorder-order transitions of gold crystals  \cite{jeon2021reversible}. These recent observations differ from the current nucleation models and call for a better theoretical insight in the crystallization pathways at the earliest stages of nucleation,  when particles start to order from the metastable fluid phase and select the emerging crystal polymorphs. 

Colloidal suspensions are suitable experimental systems to 
probe
locally heterogeneous phenomena such as
early-stage nucleation: the larger sizes and slower time scales of colloidal particles enable direct observation  of the nucleation mechanisms 
\cite{pusey1989structure,gasser2001real}.  However, even for 
hard spheres (HSs), undoubtedly 
one of the simplest colloidal model systems, the polymorph selection mechanism is yet to be revealed. 
In a HS system, the hcp crystal is metastable with respect to the fcc, but the free-energy difference between the two structures is tiny 
($\simeq 10^{-3} k_BT$ per particle) \cite{bolhuis1997entropy,noya2015entropy}. One  therefore might expect to find an approximately 50\% occurrence  of fcc- and hcp-like particles in the crystal nucleus of hard spheres.  However this prediction is not realised either in  experiments \cite{pusey1989structure,dux1997light,gasser2001real,cheng2001colloidal,sandomirski2011} nor in simulations \cite{luchnikov2002crystallization,o2003crystal,Filion2010,russo2012microscopic,leoni2021non}, which both show  a hitherto unexplained predominance of  fcc particles   in the final crystal phase.

In this Letter, we investigate, using Molecular Dynamics (MD)  simulations and particle-resolved studies of colloids, the early stages of nucleation of hard spheres in order to shed light on the selection mechanism of the crystal polymorph.  We study  the  structural transformations  in the supersaturated fluid phase that finally  lead to crystal nucleation. We find that the crystal embryo shows a preference towards fcc-like stacking, because of its 
striking similarity with local clusters present in the fluid phase. We also demonstrate that this purely geometric argument for a higher propensity to nucleate fcc is  incorporated in  thermodynamics by a lower  interfacial free energy of  fcc with respect to hcp crystals.

\begin{figure*}
\centering
\includegraphics[width=1.0\textwidth]{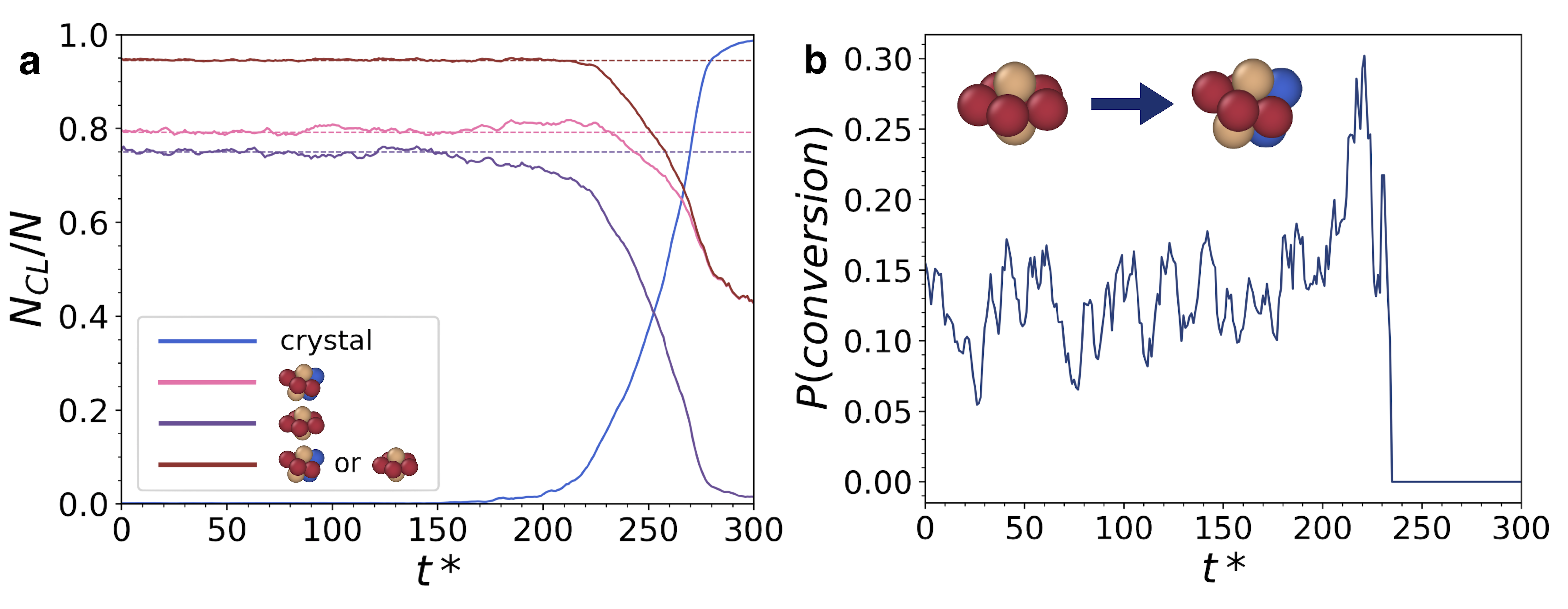}
\caption{\textbf{Behaviour of Siamese Dodecahedron (SD) and Pentagonal Bipyramid (PB) clusters during nucleation.} 
(a) Fraction of particles belonging to SD (pink), PB (purple), and combined SD or PB (red) clusters along with the fraction of solid-like particles (blue) as a function of time during an  exemplary spontaneous nucleation event. Note that a particle can be part of an SD cluster and a PB cluster at the same time, and therefore the corresponding fractions add up to a value which is higher than one. Also, a particle can be classified as crystal-like independently from whether it is also part of an SD cluster or not. The average values in the metastable fluid are shown by dashed lines. 
(b) Probability that a given PB cluster transforms into a SD cluster within a time interval of $\Delta t^* = 10$ during this nucleation event, calculated in a subcell of the system, which is centred around the centre-of-mass of the biggest crystalline cluster. We set this probability to zero when the denominator, \emph{i.e.} the number of PB clusters in the considered subcell of the system, is lower than 10 units, for poor statistics. A sketch of this conversion is shown as an inset in (b).}
\label{fig:fig1}
\end{figure*}

\subsection{Siamese Dodecahedra and Pentagonal Bipyramids} We perform MD simulations to study  crystal nucleation in a supersaturated fluid of hard spheres. We generate many nucleation events and analyse the trajectories using two different methodologies. To follow the nucleation process, we first identify the  \emph{solid-like} particles, \emph{i.e.} particles with a local solid-like (ordered)  environment, by calculating the averaged  bond order parameters \cite{lechner2008accurate}  that are based on spherical harmonics $Y_{lm}$, measuring the arrangement of the neigbours around a particle. In particular, we identify particle $i$ as solid-like  if the sixfold rotational invariant $\bar{q}_6(i) \geq 0.31$, and we colour them blue in Fig. 1c-f. We note that this classification scheme acts on a single-particle level.

To investigate the structure of the fluid,  we determine the topologies of various particle clusters present in the system using the Topological Cluster Classification (TCC) algorithm \cite{malins2013identification}. We identify local clusters of 3 up to 13 particles consisting of not only rings of three, four, and five particles  with and without additional neighbouring particles, but also  compounds of these basic clusters. In total, we distinguish 41 topological clusters.

We focus our attention on a specific cluster geometry, the Siamese Dodecahedron (SD) due to its unique behaviour in the early stages of nucleation. In addition, we consider the Pentagonal Bipyramid (PB) because of its abundance in the fluid phase and its geometric similarity with the SD cluster.
The  SD cluster  consists of  particles that occupy four out of the five vertices of a pentagonal planar ring which we refer to as \emph{ring} particles (as denoted by the red particles in Fig. 1a). The missing particle of the pentagonal ring is replaced by two  particles (denoted by the blue particles in Fig. 1a), which are shifted up and down with respect to the pentagonal planar ring. We refer to these particles as \emph{shifted} particles. Finally, two \emph{spindle} particles (gold particles in Fig. 1a) are placed on top and below the pentagonal ring. The PB cluster is  composed of \emph{ring} particles (red particles in Fig. 1b), which form a pentagonal ring with two \emph{spindle} particles (gold particles in Fig. 1b) similar to the Siamese dodecahedron.

For each particle in the system, we  calculate  the number of SD (PB) clusters that a particle belongs to. In Fig. 1c and 1d, we colour  the fluid-like particles with different shades of pink (purple), depending on the number of SD (PB) clusters they are part of, according to the scale bar on the left (right).
Even though the density of SD (PB) clusters is high throughout  the fluid, Fig. 1c and 1d show that the density of SD (PB) clusters is spatially heterogeneous. Specifically, we observe that the crystal nucleus is surrounded by a high density of SD clusters, whereas the opposite trend is found for the PB clusters as  the PB clusters are depleted near the surface of the crystal nucleus. Remarkably, the density of PB clusters seems to be anti-correlated with the density of SD clusters.

In Fig. 1e and 1f we perform the same analysis on an experimental sample, showing a similar heterogeneous structure consisting of high- and low-density regions of SD and PB clusters in the fluid phase, and a  crystal nucleus that is surrounded by a high density of SD clusters and a low density of PB clusters, in excellent agreement with our simulations. We refer the reader to the Methods Section for more details on the experiments.

\begin{figure*}
\centering
\includegraphics[width=1.0\textwidth]{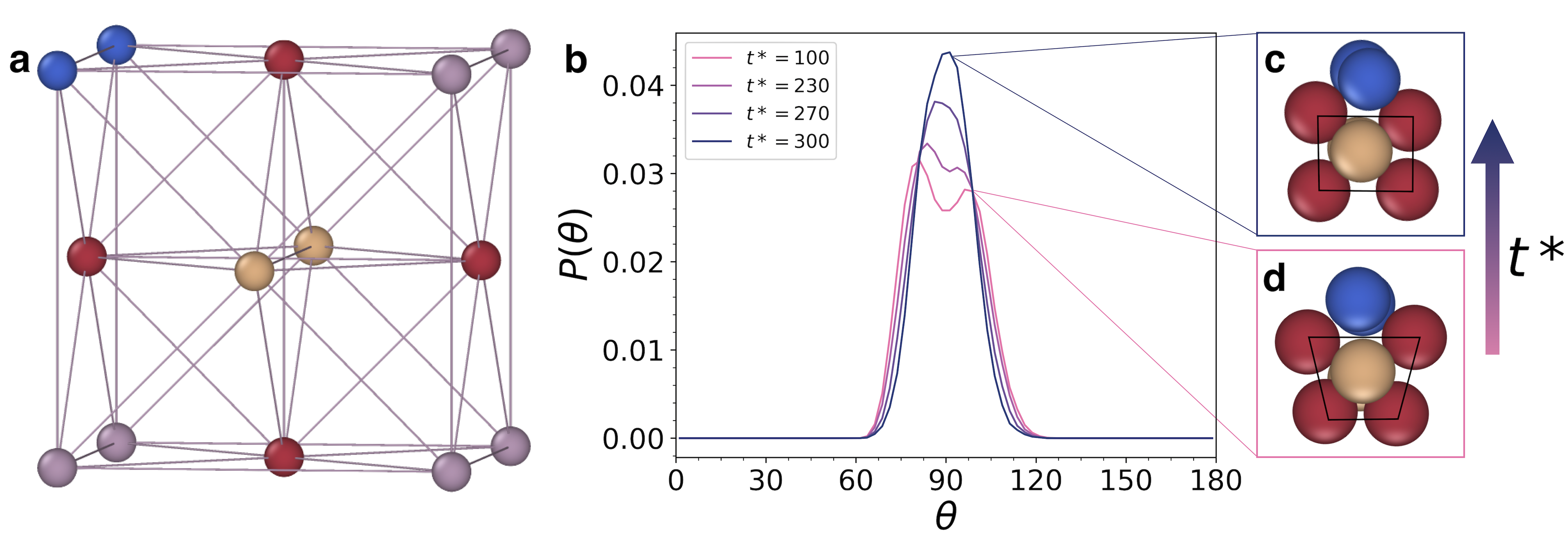}
\caption{\textbf{Transition from a Siamese Dodecahedron (SD) cluster to an fcc subunit.} (a) Arrangement of the particles in an fcc unit cell. Red, blue and golden particles correspond to an SD cluster, while the remaining  particles are coloured in lilac. (b) Probability distribution of the four angles $\theta$ of the trapezoidal arrangement of the 4 particles (red) in the pentagonal ring of all SD clusters in the system as computed at four different times during the crystallization process. The typical arrangements of particles in SD clusters after and before nucleation are shown in (c) and (d), respectively, where the black lines connecting the centres of the \emph{ring} particles help to better understand the transition.}
\label{fig:fig3}
\end{figure*}

The incompatibility of the fivefold clusters with crystalline order rationalises the depletion of PB clusters near the surface of the crystal nucleus. It is tempting to speculate that the SD clusters surrounding the crystal nucleus play a transient role in the formation of the crystal phase, which will be investigated  in more detail below.

To better understand the role of the PB and SD clusters in the crystallization mechanism of hard spheres, we plot in Fig. 2a the fraction of particles belonging to SD (pink line) and PB (purple line) clusters as a function of time during an exemplary spontaneous nucleation event  along with the fraction of crystalline particles (blue line) for comparison. Fig. 2a shows that  the fraction of crystalline particles is approximately zero in the metastable fluid phase at the beginning of this trajectory until it starts to rise when crystallization sets in. We also observe that the populations of particles in 
both the SD and PB clusters are already high before crystallization sets in, showing that the metastable fluid exhibits strong spatial correlations due to packing constraints. More surprisingly, we find an increase in the number of SD clusters during the early stages of crystallization, which decreases to a lower value at the end of the crystallization process since  SD clusters are not present in the fcc structure. In addition, the fraction of PB clusters decreases at the onset of crystallization.

To investigate the anti-correlation between SD and PB clusters, we also measure the combined fraction of particles belonging to either SD or PB clusters as a function of time (red line in Fig. 2a). The combined fraction is not only constant in the metastable fluid phase, but also shows  lesser fluctuations than the individual fractions of SD and PB clusters. More surprisingly, we observe that the combined fraction remains constant during the early stages of crystallization, thereby demonstrating that the increase in SD clusters is a consequence of a decrease of PB clusters. The constant combined fraction of SD and PB clusters and the much smaller fluctuations suggest that there is a reversible conversion between PB and SD clusters. To this end, we calculate the probability that a PB cluster transforms into an SD cluster
within a time interval $\Delta t^* =10$ by only taking into account the subcell of the system where the first nucleus appeared. In Fig. 2b, we plot the conversion rate as a function of time. We find that the rate of PB  into SD clusters is constant in the metastable fluid, and increases when  crystallization sets in.

We thus observe that the supersaturated fluid exhibits a heterogeneous structure of high- and low-density regions of PB and SD clusters with a continuous conversion between the two clusters. In addition, we find that the early stages of crystallization is signaled by a higher conversion rate of PB into SD clusters, resulting in an increased  fraction of SDs as shown in Fig. 2a.  Subsequently, the number of SDs decreases when the crystal nucleus grows further, thereby demonstrating  that the SD clusters represent an intermediate stage in the attachment of fluid-like particles to the crystal nucleus.

\subsection{The nucleation mechanism} To understand the role of  SD clusters in the fluid-solid transformation,  we note that the four particles of the pentagonal ring of an SD cluster form a trapezoidal arrangement with two acute and two obtuse angles, see Fig. 3d. Interestingly, in the case that these particles form a square arrangement (Fig. 3c), the SD cluster can be identified as a subunit of an fcc crystal as illustrated in Fig. 3a where the particles are denoted with the same colours to facilitate the comparison. Given this  topological similarity, we speculate that the attachment of fluid-like particles to the solid nucleus proceeds \emph{via} SD clusters where the four particles in the pentagonal ring transform from a trapezoidal to a square arrangement such that it becomes part of the fcc cluster.

To investigate this conjecture, we measure the distribution of the four angles of the trapezoidal arrangement of the 4 particles in the pentagonal ring of the SD clusters, at four different times during the crystallization process.
Fig. 3b shows that, in the fluid phase ($t^* = 100$), the distribution is bimodal with a peak at an angle smaller and larger than 90$^{\circ}$, representing the trapezoidal arrangement. As crystallization progresses, the distribution becomes unimodal with a single peak around 90$^{\circ}$, indicating a square pattern. 

Our results  provide strong support that the trapezoidal arrangement of the four particles in the pentagonal ring of the SD cluster transforms into a square arrangement corresponding to a subunit of the fcc crystal. This transition is also illustrated in Fig. 3c and 3d, showing two representative SD clusters after and before the transformation, respectively. The key finding of our study is that the fivefold PB clusters -- known to be inhibitors of crystal nucleation and abundant in the fluid phase -- transform into SD clusters, and that the SD cluster-mediated attachment of particles to the growing nucleus  proceeds \emph{via} a simple rearrangement of particles into fcc subunits. The rearrangement of SD clusters into hcp is less straightforward and involves an additional displacement by one of the \emph{shifted} particles (see Supplementary Information). Hence, the propensity to grow  fcc is  higher  than hcp, revealing that the polymorph selection mechanism in hard spheres is already hidden in the higher order  structure of the fluid phase.

\begin{figure*}
\centering
\includegraphics[width=0.9\textwidth]{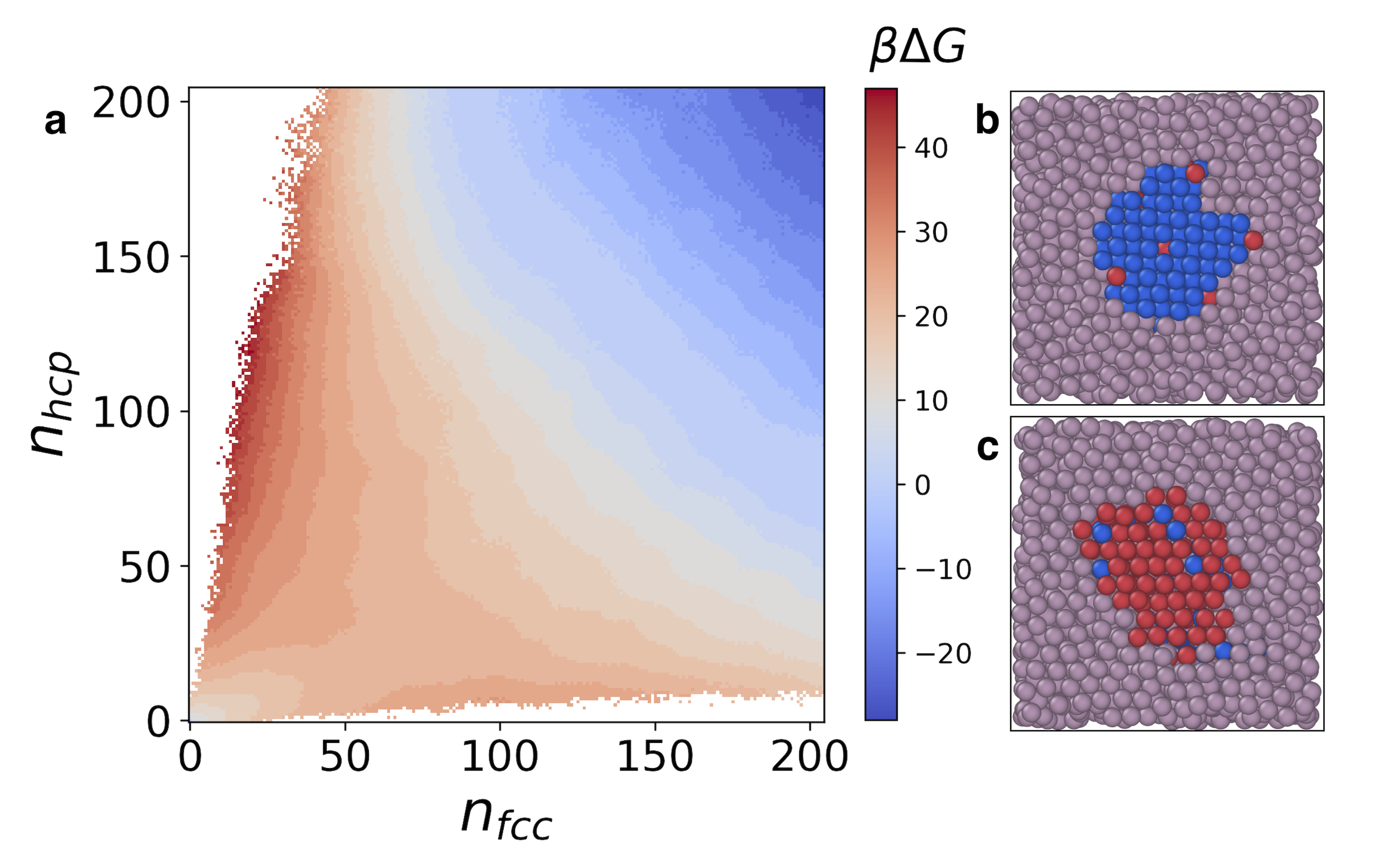}
\caption{\textbf{Thermodynamic propensity towards fcc-like particles in the early stages of crystal nucleation of hard spheres.} (a) Gibbs  free-energy barrier as a function of the number of fcc and hcp particles in the crystal nucleus as recognised by the classification scheme described in the \emph{Methods} Section. (b) A typical configuration of a nearly pure fcc crystal nucleus and of (c)  a nearly pure hcp crystal nucleus as obtained from US simulations, where fcc-like particles are coloured blue, hcp-particles are  red, and fluid-like particles are  lilac.}
\label{fig:fig4}
\end{figure*}

\subsection{The minimum free-energy pathway for nucleation} This finding begs the crucial question whether the polymorph  selection mechanism as identified here has a kinetic or thermodynamic origin.
In other words, does an fcc crystal  have a lower interfacial free-energy -- and hence a lower Gibbs free-energy barrier -- than an hcp crystal in a metastable fluid phase? To answer this question, we calculate the Gibbs free energy $\beta \Delta G(n_{fcc},n_{hcp})$ for the formation of a crystal cluster consisting of $n_{fcc}$ fcc-like particles and $n_{hcp}$ hcp-like particles using the Umbrella Sampling (US) technique, see the Methods section for the technical details. 

In Fig. 4a, we plot $\beta \Delta G$ of a crystalline nucleus composed of $n_{fcc}$ fcc-like particles and $n_{hcp}$ hcp-like particles. The lowest free-energy path on this surface shows that the crystal nucleus has an excess of fcc-like particles in the early stages of nucleation  and that the critical nucleus consists of  about $70\%$  fcc-like particles. We show two exemplary configurations of a nearly fcc-like and  hcp-like cluster in Figs. 4b and 4c, respectively, demonstrating the effectiveness of our  umbrella sampling method to bias towards nuclei with a certain composition. 

\subsection{Conclusions} In conclusion,  we unveal the crystallization and polymorph selection mechanism in a fluid of hard spheres by analysing the early stages of nucleation in MD simulations.  We show that the supersaturated fluid is highly dynamic as there is a reversible conversion between fivefold Pentagonal Bipyramid and Siamese Dodecahedron clusters. The Siamese Dodecahedra have  a stunning similarity with an fcc subunit, thereby explaining the as-of-yet unexplained higher propensity of fcc compared to hcp in hard spheres. Finally, we show that the polymorph selection mechanism has not only a geometric origin which is hidden in the higher-order correlations of  the fluid phase, but also a thermodynamic one as the lowest free-energy path proceeds \emph{via} a higher number of fcc-like particles with respect to hcp-like particles in the early stages of nucleation.  This insight suggest ways to control the nucleation pathways and the crystal polymorphs.

\section{Methods}

\subsection{MD Simulations}
In order to generate trajectories in which we observe spontaneous nucleation, we conduct MD simulations in the isothermal-isobaric (NPT) ensemble with a constant number  $N = 13500$ of nearly-hard spheres. The particles interact \emph{via} a Weeks-Chandler-Andersen (WCA) pair potential, which can straightforwardly be employed in Molecular Dynamics (MD) simulations and which reduces to the hard-sphere potential in the limit that the temperature $T \rightarrow 0$. The WCA   potential $u(r_{ij})$  reads \cite{weeks1971role} 
\begin{equation}
u \left(r_{ij}\right) = 
\begin{cases}
    4 \epsilon \left\lbrack \left(\frac{\sigma}{r_{ij}} \right)^{12} \hspace{-2mm}- \left(\frac{\sigma}{r_{ij}} \right)^6 \hspace{-2mm}+ \frac{1}{4} \right\rbrack  &r_{ij}<2^{\frac{1}{6}}\sigma\\
   0 &r_{ij}\geq2^{\frac{1}{6}}\sigma ,
\end{cases}
\label{wca}
\end{equation}
with $r_{ij}=|\mathbf{r}_i-\mathbf{r}_j|$  the centre-of-mass distance between particle $i$ and $j$, $\mathbf{r}_i$  the position of particle $i$,  $\epsilon$ the interaction strength, and   $\sigma$ the diameter of each sphere. The steepness of the repulsion between the particles can be tuned by the temperature $k_BT/\epsilon$. We set $k_BT / \epsilon = 0.025$, which has been  used extensively in previous simulation studies  to mimic hard spheres \cite{kawasaki2010formation,filion2011simulation,speck2018a,speck2018b}.

The temperature $T$ and pressure $P$ are kept constant \emph{via} the Martyna-Tobias-Klein (MTK) integrator \cite{martyna1994constant}, with the thermostat and barostat coupling constants $\tau_{\tiny T}$ = 1.0 $\tau_{\scriptsize MD}$ and $\tau_{\tiny P}$ = 1.0 $\tau_{\scriptsize MD}$, respectively, and $\tau_{\scriptsize MD} = \sigma_{\scriptsize L}\sqrt{m/\epsilon}$ is the MD time unit. The time step is set to $\Delta t = 0.004\tau_{\scriptsize MD}$, which is small enough to ensure stability of the simulations. We ran  the simulations  for 10$^{9} \tau_{\scriptsize MD}$ time steps, unless specified otherwise. The simulation box is cubic and periodic boundary conditions are applied in all directions.

We select the pressure values in a region of metastability  that allow us to observe nucleation phenomena on reasonable time scales. Specifically, the reduced pressure varies in the range $\beta P \sigma^3 \in [13.40, 16.00]$, which results in numerous spontaneous crystallization events. All MD simulations are performed using the HOOMD-blue (Highly Optimised Object-oriented Many-particle Dynamics) software \cite{Anderson2020}. 

In order to calculate an effective packing fraction for the WCA systems, we use the mapping described in Refs. \cite{filion2011simulation,speck2018a,speck2018b}, which results in each particle having an effective diameter $\sigma_\mathrm{eff} \simeq 1.097$.

\subsection{Experiments}

We used polymethyl methacrylate (PMMA) particles of diameter 2.00 mm with a polydispersity of 4.0\% as determined by static light scattering which were fluorescently labelled with Rhodamine dye. The particles were dispersed in a density matching mixture of cis decalin and cyclohexyl bormide. Tetrabutyl ammonium bromide salt was used to screen the electrostatic charges. The resulting dispersions were imaged using a Leica SP5 confocal microscope. Due to the residual electrostatic interactions, the effective hard sphere diameter is 1.02 times that of the physical diameter and thus we quote experimental values in effective packing fraction. Further details are available in \cite{taffs2013structure}.

\subsection{Bond Order Parameters}

To describe the local environment of a particle, we employ the standard  bond-orientational order parameters introduced by Steinhardt {\em et al.}  \cite{steinhardt1983bond}. We first define the complex vector $q_{lm}(i)$ for  each particle $i$
\begin{equation}
 q_{lm}(i)=\frac{1}{N_b(i)}\sum\limits_{j=1}^{N_b(i)} Y_{lm}(\theta(\mathbf{r}_{ij}),\phi(\mathbf{r}_{ij})),  
 \end{equation}
where $N_b(i)$ is the number of neighbours of particle $i$,   $Y_{lm}(\theta(\mathbf{r}_{ij}),\phi(\mathbf{r}_{ij}))$ denotes the spherical harmonics, $m \in \lbrack -l,l\rbrack$, $\theta(\mathbf{r}_{ij})$ and $\phi(\mathbf{r}_{ij})$ are the polar and azimuthal angles of the distance vector  $\mathbf{r}_{ij}=\mathbf{r}_{j}-\mathbf{r}_i$, and $\mathbf{r}_{i}$ denotes the position  of particle $i$.

The averaged $\bar{q}_{lm}(i)$ is defined as 
\begin{equation}
 \bar{q}_{lm}(i)=\frac{1}{\tilde{N}_b(i)}\sum\limits_{j=1}^{\tilde{N}_b(i)}q_{lm}(j),
\end{equation}
where $\tilde{N}_b(i)$ is the number of neighbours  including particle $i$ itself. 
The rotationally invariant  quadratic and cubic averaged bond order
parameters are defined as  
\begin{eqnarray}
 \bar{q}_{l}(i)&=&\sqrt{\frac{4\pi}{2l+1}\sum\limits_{m=-l}^{l}|\bar{q}_{lm}(i)|^2},   
\end{eqnarray}
and 
\begin{eqnarray}
&\hspace{-1cm}\bar{w}_l(i)= \mathlarger{\frac{\mathlarger{\sum}\limits_{m_1+m_2+m_3}\hspace{-5mm}\big (\begin{smallmatrix}
l & l & l\\ 
m_1 & m_2 & m_3 
\end{smallmatrix}\big ) \bar{q}_{lm_1}(i) \bar{q}_{lm_2}(i)\bar{q}_{lm_3}(i) }{\Large (\mathlarger{\sum}\limits_{m=-l}^{l}|\bar{q}_{lm}(i)|\Large )^{3/2}}}.
\end{eqnarray}
To identify  the neighbours of particle $i$ we employ the parameter-free solid-angle-based nearest-neighbour (SANN) algorithm of Van Meel \cite{van2012parameter}.  This algorithm assigns a solid angle to every potential neighbour $j$ of $i$, and defines the neighbourhood of particle $i$ to consist of the $N_b(i)$  particles nearest to $i$ for which the sum of solid angles equals 4$\pi$.

\subsection{Topological Cluster Classification}

In order to perform an analysis which is not solely based on local symmetries, we require an algorithm that is capable of successfully finding different topological clusters in a metastable fluid. To this end, we employ the Topological Cluster Classification (TCC) algorithm \cite{malins2013identification}. The bonds between particles are detected using a modified Voronoi construction method. The free parameter $f_c$, controlling the amount of asymmetry that a four-membered ring can show before being identified as two three-membered rings, is set to $0.82$.

\subsection{Umbrella Sampling Simulations}
To investigate the thermodynamic propensity towards fcc-like or hcp-like ordering during the nucleation process, we use Umbrella Sampling \cite{Torrie1977umbrella} to calculate the the nucleation barrier of a system of hard spheres at a pressure of $\beta P \sigma^3 = 17.0$. Similar to previous literature \cite{auer2001prediction,Filion2010} we identify the nucleus by using the dot product
\begin{equation}
    d_l(i,j) = \frac{ \sum_{m=-l}^{l} q_{lm}(i) q_{lm}^*(j) } { \left(\sum_{m=-l}^{l} |q_{lm}(i)|^2\right)^{1/2}\left(\sum_{m=-l}^{l} |q_{lm}(j)|^2\right)^{1/2} },
\end{equation}
with $l=6$ to define solid-like bonds as those bonds between particle pairs $(i,j)$ for which $d_6(i,j)>0.7$, and define solid-like particles as those that have at least 7 of such solid-like bonds. Particle neighbours are defined using a distance cutoff of $r_c=1.4\sigma$. The nucleus is then the largest set of solid-like particles that are connected by solid-like bonds. To disentangle fcc-like and hcp-like order, we subsequently classify solid-like particles as fcc-like and hcp-like based on their value of the Steinhardt bond order parameter $w_4$: particles with $w_4<0$ are fcc-like and those with $w_4 \geq 0$ are hcp-like \cite{lechner2008accurate}. The number of such particles are $n_{fcc}$ and $n_{hcp}$, respectively, and we use these to define the US biasing potential:
\begin{equation}
\label{eq:BiasU}
    U_b = \frac{1}{2}\lambda_{fcc} \left(n_{fcc} - n_0^{fcc}\right)^2 + \frac{1}{2}\lambda_{hcp} \left(n_{hcp} - n_0^{hcp}\right)^2,
\end{equation}
where both coupling constants $\lambda_{fcc}$ and $\lambda_{hcp}$ are set to an equal value of $\beta\lambda_{fcc}=\beta\lambda_{hcp}=0.05$. This allows us to sample the two-dimensional  Gibbs free-energy difference $\beta \Delta G(n_{fcc},n_{hcp})$ that is the nucleation barrier as a function of the number of fcc-like and hcp-like ordered particles. 

We initialise each US window $(n_0^{fcc}, n_0^{hcp})$ from a configuration with a nucleus with approximately $n_{fcc}\approx n_0^{fcc}$ and $n_{hcp}\approx n_0^{hcp}$. For very small nuclei up to $n = n_{fcc}+n_{hcp} \sim 20$ we measure the full cluster size distribution instead of only the size of the largest cluster, as the probability of multiple small nuclei appearing simultaneously can be significant.
We implement the US scheme by adding additional Monte Carlo bias moves that accept or reject trajectories based on the bias potential of \ref{eq:BiasU} on top of a hard-particle Monte Carlo (HPMC) simulation implemented using HOOMD-blue's HPMC module \cite{Anderson2020,Anderson2016}.
Bias moves are performed every MC cycle in order to also sample regions of the free-energy landscape where the gradient is large.
Finally, we reconstruct the nucleation barrier by using the Weighted Histogram Analysis Method (WHAM) \cite{Kumar1992wham}, specifically by using the algorithm provided by Ref. \cite{Grossfield2021whamcode}.

\section{Data availability}

The data associated with this research is available upon reasonable request.

\section{Code availability}

The simulation and analysis codes associated with this research are available upon reasonable request.

\bibliography{ref}

\begin{thebibliography}{10}
\expandafter\ifx\csname url\endcsname\relax
  \def\url#1{\texttt{#1}}\fi
\expandafter\ifx\csname urlprefix\endcsname\relax\def\urlprefix{URL }\fi
\providecommand{\bibinfo}[2]{#2}
\providecommand{\eprint}[2][]{\url{#2}}

\bibitem{pusey1989structure}
\bibinfo{author}{Pusey, P.} \emph{et~al.}
\newblock \bibinfo{title}{Structure of crystals of hard colloidal spheres}.
\newblock \emph{\bibinfo{journal}{Physical Review Letters}}
  \textbf{\bibinfo{volume}{63}}, \bibinfo{pages}{2753} (\bibinfo{year}{1989}).

\bibitem{palberg1997colloidal}
\bibinfo{author}{Palberg, T.}
\newblock \bibinfo{title}{Colloidal crystallization dynamics}.
\newblock \emph{\bibinfo{journal}{Current Opinion in Colloid \& Interface
  science}} \textbf{\bibinfo{volume}{2}}, \bibinfo{pages}{607--614}
  (\bibinfo{year}{1997}).

\bibitem{cheng2001colloidal}
\bibinfo{author}{Cheng, Z.}, \bibinfo{author}{Zhu, J.},
  \bibinfo{author}{Russel, W.~B.}, \bibinfo{author}{Meyer, W.~V.} \&
  \bibinfo{author}{Chaikin, P.~M.}
\newblock \bibinfo{title}{Colloidal hard-sphere crystallization kinetics in
  microgravity and normal gravity}.
\newblock \emph{\bibinfo{journal}{Applied Optics}}
  \textbf{\bibinfo{volume}{40}}, \bibinfo{pages}{4146--4151}
  (\bibinfo{year}{2001}).

\bibitem{filion2011simulation}
\bibinfo{author}{Filion, L.}, \bibinfo{author}{Ni, R.},
  \bibinfo{author}{Frenkel, D.} \& \bibinfo{author}{Dijkstra, M.}
\newblock \bibinfo{title}{Simulation of nucleation in almost hard-sphere
  colloids: The discrepancy between experiment and simulation persists}.
\newblock \emph{\bibinfo{journal}{The Journal of Chemical Physics}}
  \textbf{\bibinfo{volume}{134}}, \bibinfo{pages}{134901}
  (\bibinfo{year}{2011}).

\bibitem{russo2012microscopic}
\bibinfo{author}{Russo, J.} \& \bibinfo{author}{Tanaka, H.}
\newblock \bibinfo{title}{The microscopic pathway to crystallization in
  supercooled liquids}.
\newblock \emph{\bibinfo{journal}{Scientific Reports}}
  \textbf{\bibinfo{volume}{2}}, \bibinfo{pages}{1--8} (\bibinfo{year}{2012}).

\bibitem{sandomirski2011}
\bibinfo{author}{Sandomirski, K.}, \bibinfo{author}{Allahyarov, H., E.Loewen}
  \& \bibinfo{author}{Egelhaaf, S.~U.}
\newblock \bibinfo{title}{Heterogeneous crystallization of hard-sphere colloids
  near a wall}.
\newblock \emph{\bibinfo{journal}{Soft Matter}} \textbf{\bibinfo{volume}{7}},
  \bibinfo{pages}{8050} (\bibinfo{year}{2011}).

\bibitem{frank1952supercooling}
\bibinfo{author}{Frank, F.~C.}
\newblock \bibinfo{title}{Supercooling of liquids}.
\newblock \emph{\bibinfo{journal}{Proceedings of the Royal Society of London.
  Series A. Mathematical and Physical Sciences}}
  \textbf{\bibinfo{volume}{215}}, \bibinfo{pages}{43--46}
  (\bibinfo{year}{1952}).

\bibitem{taffs2016role}
\bibinfo{author}{Taffs, J.} \& \bibinfo{author}{Royall, C.~P.}
\newblock \bibinfo{title}{The role of fivefold symmetry in suppressing
  crystallization}.
\newblock \emph{\bibinfo{journal}{Nature Communications}}
  \textbf{\bibinfo{volume}{7}}, \bibinfo{pages}{1--7} (\bibinfo{year}{2016}).

\bibitem{sear2007nucleation}
\bibinfo{author}{Sear, R.~P.}
\newblock \bibinfo{title}{Nucleation: theory and applications to protein
  solutions and colloidal suspensions}.
\newblock \emph{\bibinfo{journal}{Journal of Physics: Condensed Matter}}
  \textbf{\bibinfo{volume}{19}}, \bibinfo{pages}{033101}
  (\bibinfo{year}{2007}).

\bibitem{palberg2014crystallization}
\bibinfo{author}{Palberg, T.}
\newblock \bibinfo{title}{Crystallization kinetics of colloidal model
  suspensions: recent achievements and new perspectives}.
\newblock \emph{\bibinfo{journal}{Journal of Physics: Condensed Matter}}
  \textbf{\bibinfo{volume}{26}}, \bibinfo{pages}{333101}
  (\bibinfo{year}{2014}).

\bibitem{ohm1995apolipoprotein}
\bibinfo{author}{Ohm, T.} \emph{et~al.}
\newblock \bibinfo{title}{Apolipoprotein e polymorphism influences not only
  cerebral senile plaque load but also alzheimer-type neurofibrillary tangle
  formation}.
\newblock \emph{\bibinfo{journal}{Neuroscience}} \textbf{\bibinfo{volume}{66}},
  \bibinfo{pages}{583--587} (\bibinfo{year}{1995}).

\bibitem{bauer2001ritonavir}
\bibinfo{author}{Bauer, J.} \emph{et~al.}
\newblock \bibinfo{title}{Ritonavir: an extraordinary example of conformational
  polymorphism}.
\newblock \emph{\bibinfo{journal}{Pharmaceutical Research}}
  \textbf{\bibinfo{volume}{18}}, \bibinfo{pages}{859--866}
  (\bibinfo{year}{2001}).

\bibitem{van2018molecular}
\bibinfo{author}{Van~Driessche, A.~E.} \emph{et~al.}
\newblock \bibinfo{title}{Molecular nucleation mechanisms and control
  strategies for crystal polymorph selection}.
\newblock \emph{\bibinfo{journal}{Nature}} \textbf{\bibinfo{volume}{556}},
  \bibinfo{pages}{89--94} (\bibinfo{year}{2018}).

\bibitem{xing2019atomistic}
\bibinfo{author}{Xing, J.}, \bibinfo{author}{Schweighauser, L.},
  \bibinfo{author}{Okada, S.}, \bibinfo{author}{Harano, K.} \&
  \bibinfo{author}{Nakamura, E.}
\newblock \bibinfo{title}{Atomistic structures and dynamics of prenucleation
  clusters in mof-2 and mof-5 syntheses}.
\newblock \emph{\bibinfo{journal}{Nature communications}}
  \textbf{\bibinfo{volume}{10}}, \bibinfo{pages}{1--9} (\bibinfo{year}{2019}).

\bibitem{zhou2019observing}
\bibinfo{author}{Zhou, J.} \emph{et~al.}
\newblock \bibinfo{title}{Observing crystal nucleation in four dimensions using
  atomic electron tomography}.
\newblock \emph{\bibinfo{journal}{Nature}} \textbf{\bibinfo{volume}{570}},
  \bibinfo{pages}{500--503} (\bibinfo{year}{2019}).

\bibitem{houben2020mechanism}
\bibinfo{author}{Houben, L.}, \bibinfo{author}{Weissman, H.},
  \bibinfo{author}{Wolf, S.~G.} \& \bibinfo{author}{Rybtchinski, B.}
\newblock \bibinfo{title}{A mechanism of ferritin crystallization revealed by
  cryo-stem tomography}.
\newblock \emph{\bibinfo{journal}{Nature}} \textbf{\bibinfo{volume}{579}},
  \bibinfo{pages}{540--543} (\bibinfo{year}{2020}).

\bibitem{nakamuro2021capturing}
\bibinfo{author}{Nakamuro, T.}, \bibinfo{author}{Sakakibara, M.},
  \bibinfo{author}{Nada, H.}, \bibinfo{author}{Harano, K.} \&
  \bibinfo{author}{Nakamura, E.}
\newblock \bibinfo{title}{Capturing the moment of emergence of crystal nucleus
  from disorder}.
\newblock \emph{\bibinfo{journal}{Journal of the American Chemical Society}}
  \textbf{\bibinfo{volume}{143}}, \bibinfo{pages}{1763--1767}
  (\bibinfo{year}{2021}).

\bibitem{jeon2021reversible}
\bibinfo{author}{Jeon, S.} \emph{et~al.}
\newblock \bibinfo{title}{Reversible disorder-order transitions in atomic
  crystal nucleation}.
\newblock \emph{\bibinfo{journal}{Science}} \textbf{\bibinfo{volume}{371}},
  \bibinfo{pages}{498--503} (\bibinfo{year}{2021}).

\bibitem{gasser2001real}
\bibinfo{author}{Gasser, U.}, \bibinfo{author}{Weeks, E.~R.},
  \bibinfo{author}{Schofield, A.}, \bibinfo{author}{Pusey, P.} \&
  \bibinfo{author}{Weitz, D.}
\newblock \bibinfo{title}{Real-space imaging of nucleation and growth in
  colloidal crystallization}.
\newblock \emph{\bibinfo{journal}{Science}} \textbf{\bibinfo{volume}{292}},
  \bibinfo{pages}{258--262} (\bibinfo{year}{2001}).

\bibitem{bolhuis1997entropy}
\bibinfo{author}{Bolhuis, P.~G.}, \bibinfo{author}{Frenkel, D.},
  \bibinfo{author}{Mau, S.-C.} \& \bibinfo{author}{Huse, D.~A.}
\newblock \bibinfo{title}{Entropy difference between crystal phases}.
\newblock \emph{\bibinfo{journal}{Nature}} \textbf{\bibinfo{volume}{388}},
  \bibinfo{pages}{235--236} (\bibinfo{year}{1997}).

\bibitem{noya2015entropy}
\bibinfo{author}{Noya, E.~G.} \& \bibinfo{author}{Almarza, N.~G.}
\newblock \bibinfo{title}{Entropy of hard spheres in the close-packing limit}.
\newblock \emph{\bibinfo{journal}{Molecular Physics}}
  \textbf{\bibinfo{volume}{113}}, \bibinfo{pages}{1061--1068}
  (\bibinfo{year}{2015}).

\bibitem{dux1997light}
\bibinfo{author}{Dux, C.} \& \bibinfo{author}{Versmold, H.}
\newblock \bibinfo{title}{Light diffraction from shear ordered colloidal
  dispersions}.
\newblock \emph{\bibinfo{journal}{Physical Review Letters}}
  \textbf{\bibinfo{volume}{78}}, \bibinfo{pages}{1811} (\bibinfo{year}{1997}).

\bibitem{luchnikov2002crystallization}
\bibinfo{author}{Luchnikov, V.}, \bibinfo{author}{Gervois, A.},
  \bibinfo{author}{Richard, P.}, \bibinfo{author}{Oger, L.} \&
  \bibinfo{author}{Troadec, J.}
\newblock \bibinfo{title}{Crystallization of dense hard sphere packings:
  Competition of hcp and fcc close order}.
\newblock \emph{\bibinfo{journal}{Journal of Molecular Liquids}}
  \textbf{\bibinfo{volume}{96}}, \bibinfo{pages}{185--194}
  (\bibinfo{year}{2002}).

\bibitem{o2003crystal}
\bibinfo{author}{O’malley, B.} \& \bibinfo{author}{Snook, I.}
\newblock \bibinfo{title}{Crystal nucleation in the hard sphere system}.
\newblock \emph{\bibinfo{journal}{Physical Review Letters}}
  \textbf{\bibinfo{volume}{90}}, \bibinfo{pages}{085702}
  (\bibinfo{year}{2003}).

\bibitem{Filion2010}
\bibinfo{author}{Filion, L.}, \bibinfo{author}{Hermes, M.},
  \bibinfo{author}{Ni, R.} \& \bibinfo{author}{Dijkstra, M.}
\newblock \bibinfo{title}{{Crystal nucleation of hard spheres using molecular
  dynamics, umbrella sampling, and forward flux sampling: A comparison of
  simulation techniques}}.
\newblock \emph{\bibinfo{journal}{The Journal of Chemical Physics}}
  \textbf{\bibinfo{volume}{133}}, \bibinfo{pages}{244115}
  (\bibinfo{year}{2010}).

\bibitem{leoni2021non}
\bibinfo{author}{Leoni, F.} \& \bibinfo{author}{Russo, J.}
\newblock \bibinfo{title}{Non-classical nucleation pathways in
  stacking-disordered crystals}.
\newblock \emph{\bibinfo{journal}{arXiv preprint arXiv:2105.05506}}
  (\bibinfo{year}{2021}).

\bibitem{lechner2008accurate}
\bibinfo{author}{Lechner, W.} \& \bibinfo{author}{Dellago, C.}
\newblock \bibinfo{title}{Accurate determination of crystal structures based on
  averaged local bond order parameters}.
\newblock \emph{\bibinfo{journal}{The Journal of Chemical Physics}}
  \textbf{\bibinfo{volume}{129}}, \bibinfo{pages}{114707}
  (\bibinfo{year}{2008}).

\bibitem{malins2013identification}
\bibinfo{author}{Malins, A.}, \bibinfo{author}{Williams, S.~R.},
  \bibinfo{author}{Eggers, J.} \& \bibinfo{author}{Royall, C.~P.}
\newblock \bibinfo{title}{Identification of structure in condensed matter with
  the topological cluster classification}.
\newblock \emph{\bibinfo{journal}{The Journal of Chemical Physics}}
  \textbf{\bibinfo{volume}{139}}, \bibinfo{pages}{234506}
  (\bibinfo{year}{2013}).

\bibitem{weeks1971role}
\bibinfo{author}{Weeks, J.~D.}, \bibinfo{author}{Chandler, D.} \&
  \bibinfo{author}{Andersen, H.~C.}
\newblock \bibinfo{title}{Role of repulsive forces in determining the
  equilibrium structure of simple liquids}.
\newblock \emph{\bibinfo{journal}{The Journal of Chemical Physics}}
  \textbf{\bibinfo{volume}{54}}, \bibinfo{pages}{5237--5247}
  (\bibinfo{year}{1971}).

\bibitem{kawasaki2010formation}
\bibinfo{author}{Kawasaki, T.} \& \bibinfo{author}{Tanaka, H.}
\newblock \bibinfo{title}{Formation of a crystal nucleus from liquid}.
\newblock \emph{\bibinfo{journal}{Proceedings of the National Academy of
  Sciences}} \textbf{\bibinfo{volume}{107}}, \bibinfo{pages}{14036--14041}
  (\bibinfo{year}{2010}).

\bibitem{speck2018a}
\bibinfo{author}{Richard, D.} \& \bibinfo{author}{Speck, T.}
\newblock \bibinfo{title}{Crystallization of hard spheres revisited. i.
  extracting kinetics and free energy landscape from forward flux sampling}.
\newblock \emph{\bibinfo{journal}{The Journal of Chemical Physics}}
  \textbf{\bibinfo{volume}{148}}, \bibinfo{pages}{124110}
  (\bibinfo{year}{2018}).

\bibitem{speck2018b}
\bibinfo{author}{Richard, D.} \& \bibinfo{author}{Speck, T.}
\newblock \bibinfo{title}{Crystallization of hard spheres revisited. ii.
  thermodynamic modeling, nucleation work, and the surface of tension}.
\newblock \emph{\bibinfo{journal}{The Journal of Chemical Physics}}
  \textbf{\bibinfo{volume}{148}}, \bibinfo{pages}{224102}
  (\bibinfo{year}{2018}).

\bibitem{martyna1994constant}
\bibinfo{author}{Martyna, G.~J.}, \bibinfo{author}{Tobias, D.~J.} \&
  \bibinfo{author}{Klein, M.~L.}
\newblock \bibinfo{title}{Constant pressure molecular dynamics algorithms}.
\newblock \emph{\bibinfo{journal}{The Journal of Chemical Physics}}
  \textbf{\bibinfo{volume}{101}}, \bibinfo{pages}{4177--4189}
  (\bibinfo{year}{1994}).

\bibitem{Anderson2020}
\bibinfo{author}{Anderson, J.~A.}, \bibinfo{author}{Glaser, J.} \&
  \bibinfo{author}{Glotzer, S.~C.}
\newblock \bibinfo{title}{{HOOMD-blue: A Python package for high-performance
  molecular dynamics and hard particle Monte Carlo simulations}}.
\newblock \emph{\bibinfo{journal}{Computational Materials Science}}
  \textbf{\bibinfo{volume}{173}}, \bibinfo{pages}{109363}
  (\bibinfo{year}{2020}).
\newblock \eprint{1308.5587}.

\bibitem{taffs2013structure}
\bibinfo{author}{Taffs, J.}, \bibinfo{author}{Williams, S.~R.},
  \bibinfo{author}{Tanaka, H.} \& \bibinfo{author}{Royall, C.~P.}
\newblock \bibinfo{title}{Structure and kinetics in the freezing of nearly hard
  spheres}.
\newblock \emph{\bibinfo{journal}{Soft Matter}} \textbf{\bibinfo{volume}{9}},
  \bibinfo{pages}{297--305} (\bibinfo{year}{2013}).

\bibitem{steinhardt1983bond}
\bibinfo{author}{Steinhardt, P.~J.}, \bibinfo{author}{Nelson, D.~R.} \&
  \bibinfo{author}{Ronchetti, M.}
\newblock \bibinfo{title}{Bond-orientational order in liquids and glasses}.
\newblock \emph{\bibinfo{journal}{Physical Review B}}
  \textbf{\bibinfo{volume}{28}}, \bibinfo{pages}{784} (\bibinfo{year}{1983}).

\bibitem{van2012parameter}
\bibinfo{author}{van Meel, J.~A.}, \bibinfo{author}{Filion, L.},
  \bibinfo{author}{Valeriani, C.} \& \bibinfo{author}{Frenkel, D.}
\newblock \bibinfo{title}{A parameter-free, solid-angle based, nearest-neighbor
  algorithm}.
\newblock \emph{\bibinfo{journal}{The Journal of Chemical Physics}}
  \textbf{\bibinfo{volume}{136}}, \bibinfo{pages}{234107}
  (\bibinfo{year}{2012}).

\bibitem{Torrie1977umbrella}
\bibinfo{author}{Torrie, G.} \& \bibinfo{author}{Valleau, J.}
\newblock \bibinfo{title}{{Nonphysical sampling distributions in Monte Carlo
  free-energy estimation: Umbrella sampling}}.
\newblock \emph{\bibinfo{journal}{Journal of Computational Physics}}
  \textbf{\bibinfo{volume}{23}}, \bibinfo{pages}{187--199}
  (\bibinfo{year}{1977}).

\bibitem{auer2001prediction}
\bibinfo{author}{Auer, S.} \& \bibinfo{author}{Frenkel, D.}
\newblock \bibinfo{title}{Prediction of absolute crystal-nucleation rate in
  hard-sphere colloids}.
\newblock \emph{\bibinfo{journal}{Nature}} \textbf{\bibinfo{volume}{409}},
  \bibinfo{pages}{1020--1023} (\bibinfo{year}{2001}).

\bibitem{Anderson2016}
\bibinfo{author}{Anderson, J.~A.}, \bibinfo{author}{{Eric Irrgang}, M.} \&
  \bibinfo{author}{Glotzer, S.~C.}
\newblock \bibinfo{title}{{Scalable Metropolis Monte Carlo for simulation of
  hard shapes}}.
\newblock \emph{\bibinfo{journal}{Computer Physics Communications}}
  \textbf{\bibinfo{volume}{204}}, \bibinfo{pages}{21--30}
  (\bibinfo{year}{2016}).
\newblock \eprint{1509.04692}.

\bibitem{Kumar1992wham}
\bibinfo{author}{Kumar, S.}, \bibinfo{author}{Rosenberg, J.~M.},
  \bibinfo{author}{Bouzida, D.}, \bibinfo{author}{Swendsen, R.~H.} \&
  \bibinfo{author}{Kollman, P.~A.}
\newblock \bibinfo{title}{{THE weighted histogram analysis method for
  free-energy calculations on biomolecules. I. The method}}.
\newblock \emph{\bibinfo{journal}{Journal of Computational Chemistry}}
  \textbf{\bibinfo{volume}{13}}, \bibinfo{pages}{1011--1021}
  (\bibinfo{year}{1992}).

\bibitem{Grossfield2021whamcode}
\bibinfo{author}{Grossfield, A.}
\newblock \bibinfo{title}{{WHAM: the weighted histogram analysis method}}.
\newblock
  \urlprefix\url{http://membrane.urmc.rochester.edu/wordpress/?page{\_}id=126}.

\end{thebibliography}


\begin{thebibliography}{}
\expandafter\ifx\csname url\endcsname\relax
  \def\url#1{\texttt{#1}}\fi
\expandafter\ifx\csname urlprefix\endcsname\relax\def\urlprefix{URL }\fi
\providecommand{\bibinfo}[2]{#2}
\providecommand{\eprint}[2][]{\url{#2}}

\end{thebibliography}

\section{Acknowledgements}
The authors are grateful to Roland Roth for his intimate knowledge of complex shapes. G. M. C. and M. D. acknowledge financial support from the NWO program Data-driven science for smart and sustainable energy research (Project number: 16DDS003).

\section{Author Contributions}
G.M.C. and M.D. initiated the project. G.M.C. performed the MD simulations, and the BOP and TCC analysis. US simulations were performed by R.v.D., while experiments were carried out by C.P.R. All authors co-wrote the manuscript, and discussed the text and interpretation of the results.

\end{document}



\makeatletter
\def\fnum@figure{\figurename\nobreakspace\textbf{\thefigure}}
\makeatother

\renewcommand{\theequation}{S\arabic{equation}}
\renewcommand{\thefigure}{S\arabic{figure}}
\renewcommand{\thesection}{}

 \renewcommand{\bibsection}{\section{References}}

\titleformat{\section}[display]
  {\centering\normalfont\scshape \bfseries}{
  \MakeUppercase{#1}}{0em}{}
  
\titlespacing\section{0pt}{20pt plus 4pt minus 2pt}{0pt plus 2pt minus 2pt}

\title{Supplementary Materials for\\ Crystal polymorph selection mechanism of hard spheres hidden in the fluid}

\author{Gabriele M. Coli$^1$,  Robin van Damme$^1$, C. Patrick Royall$^{2,3,4}$, Marjolein Dijkstra$^1$}
\affiliation{$^1$Soft Condensed Matter, Debye Institute of Nanomaterials Science, Utrecht University, Princetonplein 1, 3584 CC Utrecht, Netherlands \\
$^2$Gulliver UMR CNRS 7083, ESPCI Paris, Universit\'{e} PSL, 75005 Paris, France \\
$^3$H.H. Wills Physics Laboratory, Tyndall Avenue, Bristol, BS8 1TL, UK \\
$^4$ School of Chemistry, University of Bristol, Cantock's Close, Bristol, BS8 1TS, UK \\}

\maketitle




\onecolumngrid
\vspace{2cm} 
\begin{center}
{\bf \MakeUppercase{Contents}}
\end{center}
\vspace{-1cm}
\tableofcontents
\vspace{\columnsep}

\null
\clearpage

\section{Supplementary note: Crystal growth}

During the early stages of nucleation of hard spheres, we observe that both in simulations and experiments the crystal nucleus  is surrounded by a high density of Siamese Dodecahedron (SD) clusters and a low density of Pentagonal Bipyramid (PB) clusters.

Here, we show that this scenario persists even during crystal growth when the size of the nucleus is post-critical. In  Fig. S1a and S1b, we show a post-critical nucleus as obtained from MD simulations. In Fig. S1c and S1d, we show the same analysis for an experimental configuration. The solid-like particles are coloured blue. The fluid-like particles are coloured with different shades of pink (purple), depending on the number of SD (PB) clusters they are part of, according to the scale bar on the left (right). 

The fact that the number of SD clusters around the crystal nucleus is particularly high throughout the   crystal growth process is further evidence that these clusters play a transient role in the transformation of the disordered fluid  to the ordered crystal phase.  

\bigskip

\begin{figure*}[h]
\centering
\includegraphics[width=1.0\textwidth]{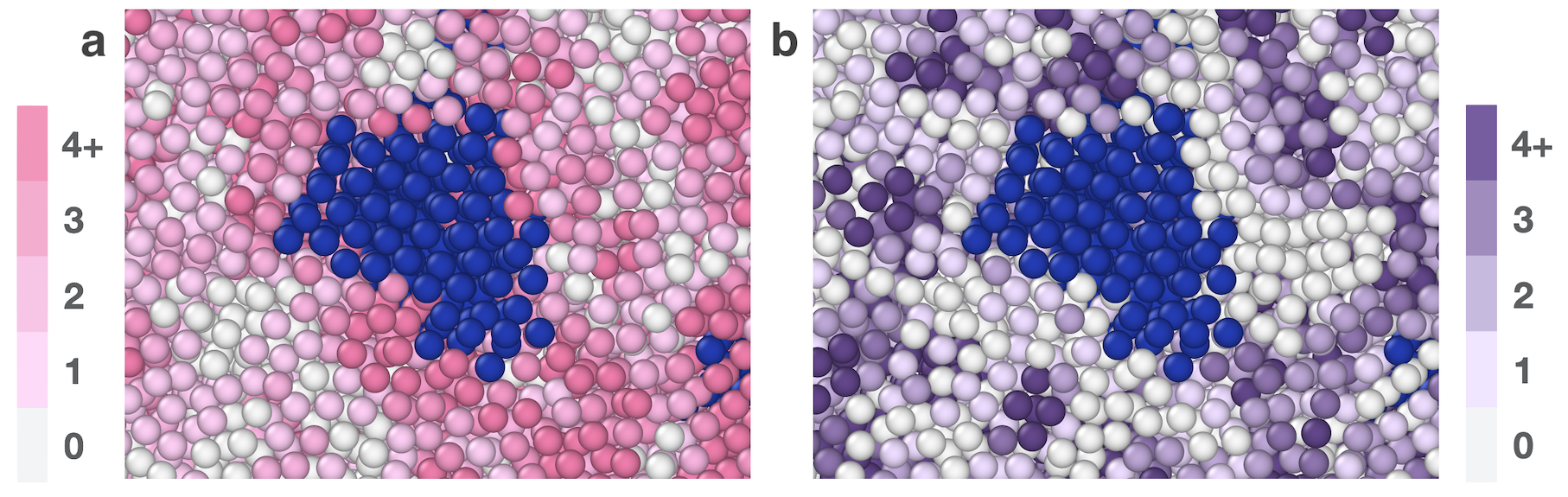}
\caption{\textbf{Typical configuration in the crystal growth regime of hard spheres.}  (a-b) Cut-through image during  crystal growth as obtained from MD simulations. Crystal-like particles are coloured blue, while fluid-like particles are coloured according to the scale bar on the left (a) or right (b) depending on the number of SD (a) or PB (b) clusters each particle belongs to.}
\label{fig:si_fig2}
\end{figure*}

\section{Supplementary note: Subcell Analysis}

To obtain a better resolution on the behaviour of SD clusters during nucleation, we divide the simulation box into 64 cubic subcells of equal  volume. 
For each subcell we compute the fraction of solid-like particles $n_x$, as well as the fraction of particles belonging to SD clusters $n_{SD}$, and to PB clusters  $n_{PB}$. Subsequently, we measure the correlation between these number fractions by computing the Pearson correlation coefficient  during the entire crystallization trajectory. The Pearson correlation coefficient $r(x,y)$ between variables $x$ and $y$ reads
\begin{equation}
r(x,y) = \frac{\sum_{i=1}^n (x_i - \bar{x}) (y_i - \bar{y})}{\sqrt{\sum_{i=1}^n (x_i - \bar{x})^2} \sqrt{\sum_{i=1}^n (y_i - \bar{y})^2}}, 
\end{equation}
where $\bar{x}$ and $\bar{y}$ represent the average values of  variable $x$ and $y$, respectively.

In Fig. S2a we plot $r(n_{SD},n_x)$ as a function of time. We observe two distinct and noteworthy features. First, at the start of the nucleation process, $n_x$ and $n_{SD}$ reach a  positive maximum correlation value, demonstrating that a subcell with  a high fraction of crystal-like particles also shows a high fraction of SD clusters.  This feature is made explicit in Fig. S2b, where each point correspond to a specific subcell of the system at the peak of the Pearson correlation between $n_{SD}$ and $n_x$. The second observation we make, is that during the crystal growth stage, the correlation suddenly drops to negative values, indicating that the crystal structure formed in the nucleus is incompatible with the presence of SD clusters.

\begin{figure*}
\centering
\includegraphics[width=1.0\textwidth]{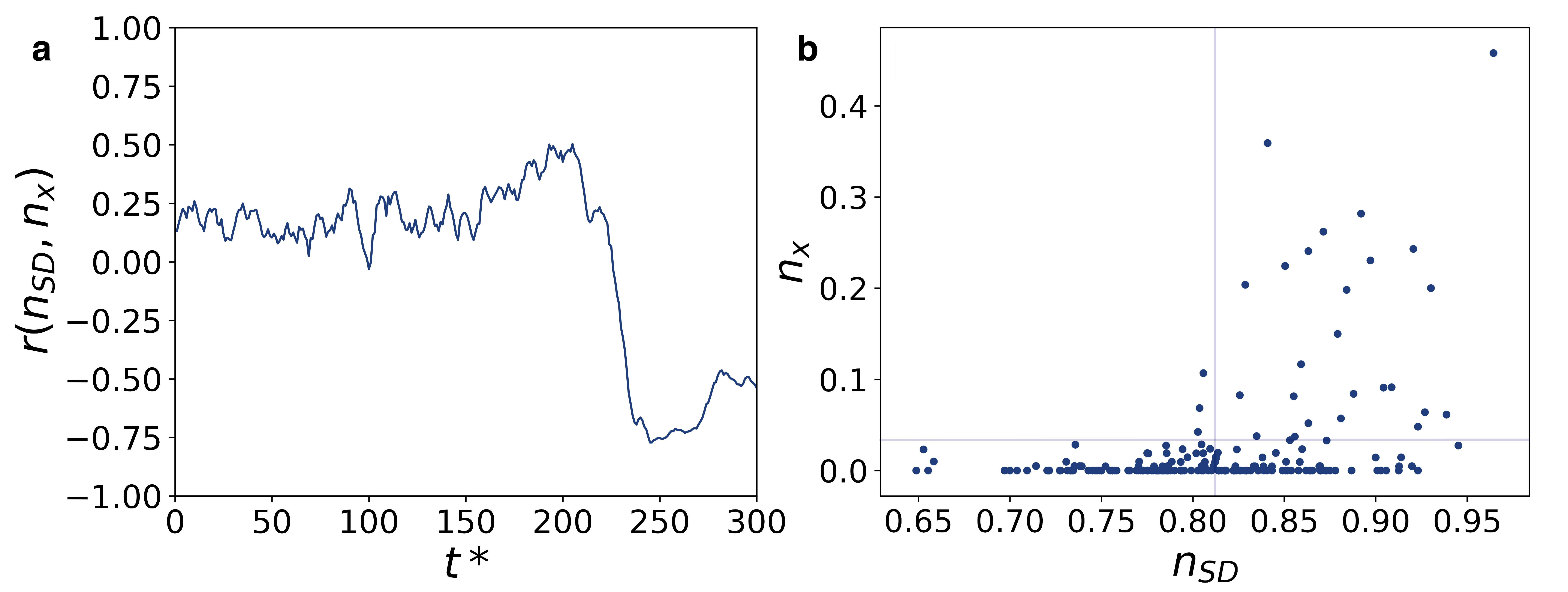}
\caption{\textbf{Subcell analysis} (a) Pearson correlation $r(n_{SD},n_x)$ as a function of time $t^*$, as defined in the main text. (b) Fraction of crystal-like particles $n_x$ as a function of the fraction of particles belonging to at least one SD cluster $n_{SD}$, calculated for each subcell in the system. The data points are from three snapshots obtained from  independent MD simulations, which  correspond to the maximum of the Pearson correlation function $r(n_{SD},n_x)$. The vertical (horizontal) line indicates the average value of $n_{SD}$ ($n_x$).}
\label{fig:si_fig2}
\end{figure*}

\section{Supplementary note: More on Bond Order Parameters}

\subsection{Distinction between ring, shifted, and spindle particles}

The analysis conducted in the main text demonstrates that SD clusters play a transient role in the attachment of  fluid particles to a crystal nucleus. To investigate this transformation further, we calculate the bond order parameters (BOPs) for the particles belonging to SD clusters. In Fig. S3a we show the BOP values for the particles composing the SD clusters in  the $\bar{q}_4 - \bar{q}_6$ plane for a typical configuration in the early stage of nucleation along with the BOPs of a typical fluid and crystal configuration for comparison. Even though a large fraction of the SD particles (bright pink points) shows a higher than average degree of fourfold and sixfold symmetry, the large spread in $\bar{q}_4$ and $\bar{q}_6$ values  show that there is no clear correlation between the SD clusters and their BOP values in the fluid phase. 

In  order to clarify the transition process from a PB cluster to an SD cluster, we divided  in the main text the particles composing an SD cluster into several categories -- \emph{ring}, \emph{shifted}, and \emph{spindle} particles. This  classification is not only useful in describing the topology of the SD clusters, but also reveals additional information regarding the nucleation mechanism. By computing the probability distribution of $\bar{q}_6$  for the three different particle types of the SD clusters at the onset of  crystallization,  we find that the \emph{shifted} particles show slightly higher $\bar{q}_6$ values (blue curve in Fig. S3b), suggesting that the crystallization process is largely initiated by the \emph{shifted}  particles.

In the main text we described how the  particles re-arrange during the transformation of  PB clusters into SD clusters. To be more specific, we showed that the  transformation is initiated by the appearance of  two \emph{shifted} particles. Recalling that the disappearance of PB clusters and the subsequent excess of SD clusters enables the start of the crystallization phenomenon, it is  to be expected that the \emph{shifted}  particles of the SD cluster are indeed the first to crystallize.

\begin{figure*}
\centering
\includegraphics[width=1.0\textwidth]{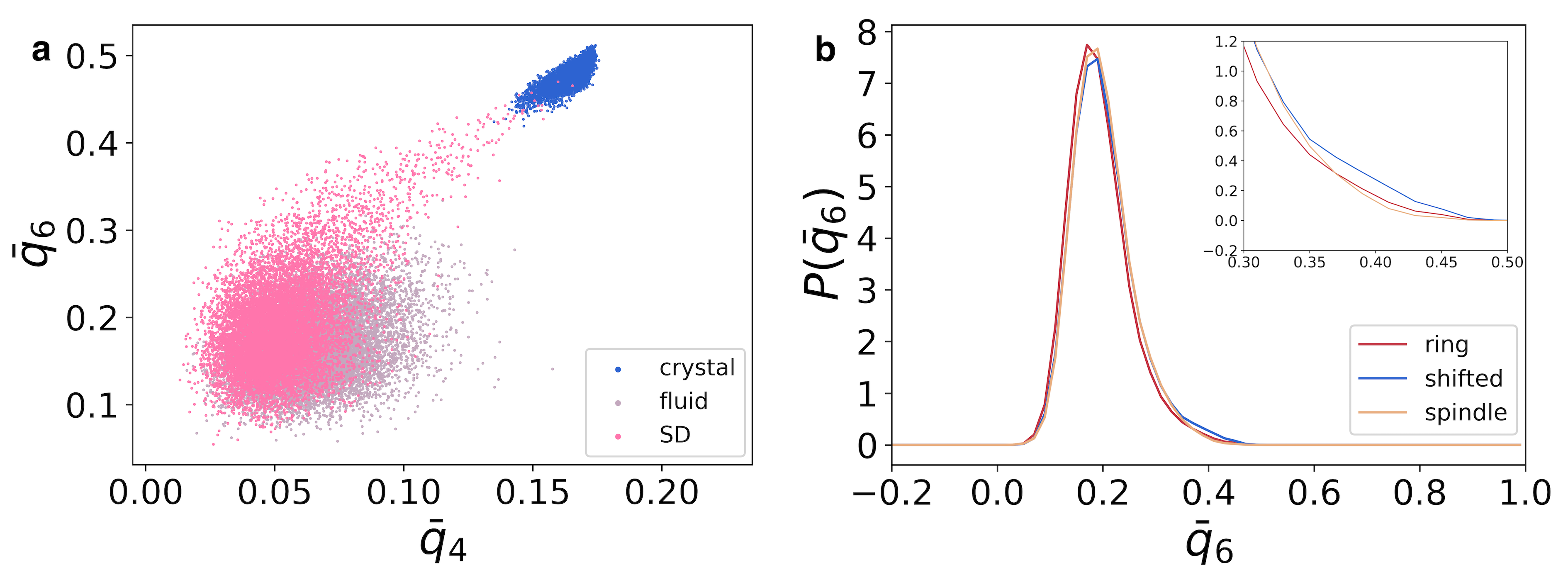}
\caption{\textbf{Correlation between bond order parameters (BOPs) and Siamese Dodecahedron (SD) clusters} (a) Projection of the BOP values of particles belonging to SD clusters in the early stages of nucleation on the $\bar{q}_4 - \bar{q}_6$ plane. Particles belonging to SD clusters show BOP values that are very similar to the ones of the fluid phase, which are therefore not useful in describing the behaviour of the SD particles. (b) Probability distribution $P(\bar{q}_6)$ of the \emph{ring}, \emph{shifted}, and \emph{spindle} particles using the same configuration as in (a). The inset shows that the \emph{shifted} particles show slightly higher $\bar{q}_6$ values, revealing that  the nucleation process is largely initiated by the  \emph{shifted} particles.}
\label{fig:si_fig3}
\end{figure*}

\subsection{Polymorph detection with different classification schemes}

The results presented in this work are all based on a combined use of bond order parameters (BOPs) and a Topological Cluster Classification (TCC) analysis. In particular, when using BOPs there are several choices to make, which affect the classification of the particles. These choices are related to the identification of the local neighbourhood of a particle, to the distinction of fluid-like and solid-like particles, and to the further distinction of the different crystal polymorphs. In this section, we select different criteria for all these sub-tasks, and check the robustness of the classification outcome. Specifically, we check that all methods record a predominance of  fcc-like  with respect to hcp-like ordering in the growing nucleus.

To this end, we start by using a total of  three different techniques to find the local neighbourhood of a particle. The first two are based on a simple cutoff radius equal to $1.4\sigma$ and $1.5\sigma$ with $\sigma$  the diameter of the particles, while the third is based on the solid-angle nearest-neighbour (SANN) algorithm. 

Furthermore, in order to classify particles as solid-like or fluid-like, we use two different criteria. One is implemented \emph{via} the criterion $\bar{q}_6 > 0.31$, as in the main text, while the second is based on dot-products of the $q_{6m}$, as implemented in the Umbrella Sampling calculations (see Methods).

Finally, in order to distinguish the different crystal polymorphs, we again use different strategies. In the first method, we employ the $w_4$ value, following the scheme proposed in our Umbrella Sampling calculations (see Methods). Alternatively, we can also use  the averaged bond order parameter $\bar{w}_4$.

Using all possible combinations for detecting the local environment, distinguishing between solid-like and fluid-like particles, and classifying different polymorphs, we obtain a total of $12$ classification algorithms. We  use all of these to analyse the spontaneous nucleation trajectory shown in the main text and compute the fraction of fcc-like and hcp-like particles in the largest crystal nucleus during a nucleation trajectory.  In Fig. S4a (Fig. S4b), we show the ratio of fcc-like (hcp-like) particles computed \emph{via} each classification scheme. We note that the first part of the nucleation event is  noisy  as the denominator, \emph{i.e.} the number of particles belonging to the main cluster, is very small. 

\begin{figure*}
\centering
\includegraphics[width=1.0\textwidth]{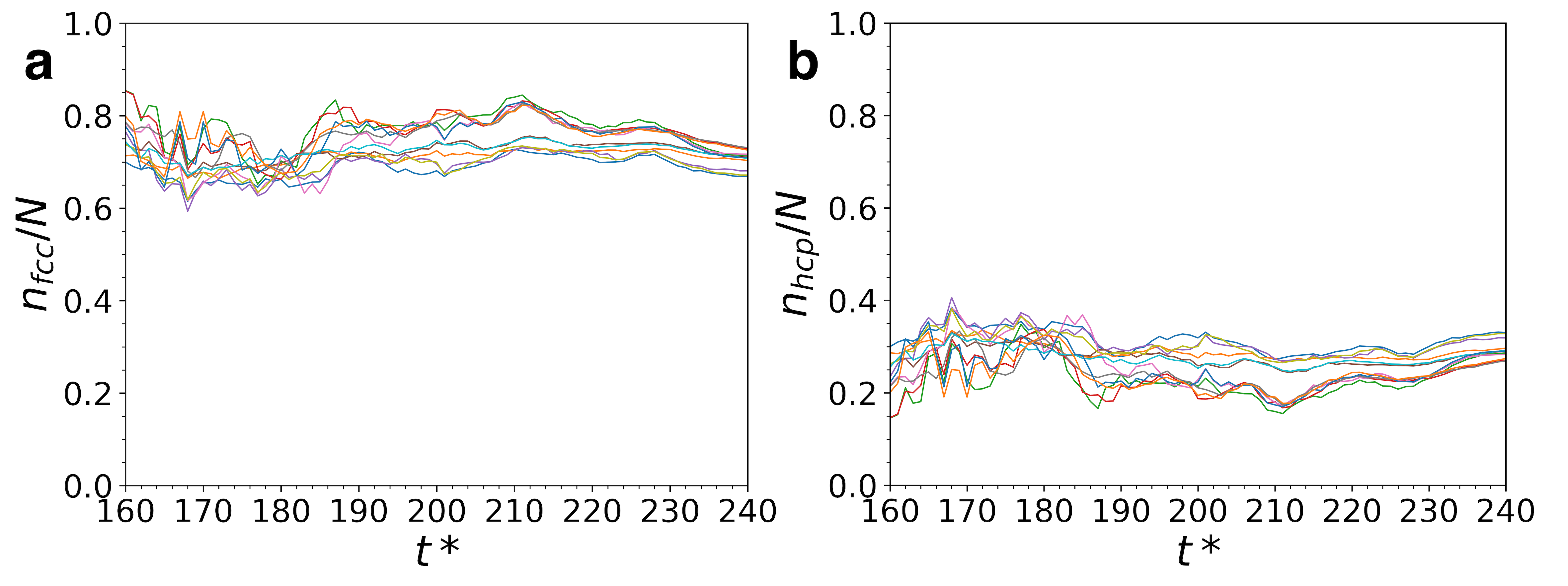}
\caption{\textbf{Crystal polymorphs in the growing crystalline cluster}  Fraction of (a) fcc-like $n_{\textrm fcc}/N$ and (b) hcp-like $n_{\textrm hcp}/N$ particles in the growing nucleus consisting of $N$ particles as identified by  $12$ different classification schemes described in the text.}
\label{fig:si_fig4}
\end{figure*}

We clearly observe that our results are robust with respect to the choice of classification scheme, thereby providing confidence that the crystal nuclei are predominately composed of (60\%-80\%) fcc-like particles.

\section{From Siamese Dodecahedra to hcp}

In Fig. 3 of the main text, we show that particles arranged in an SD cluster have a high propensity to transform to  fcc  due to the topological similarity between the SD cluster and a subunit of  fcc. In particular, the transition between the SD cluster into an fcc subunit proceeds  by changing the  trapezoidal arrangement of the four \emph{ring} particles into a square arrangement. 
Here, we show that such a simple transformation does not hold for the transition from an SD cluster to  hcp, which involves an additional displacement by one of the \emph{shifted} particles.

\begin{figure*}
\centering
\includegraphics[width=0.6\textwidth]{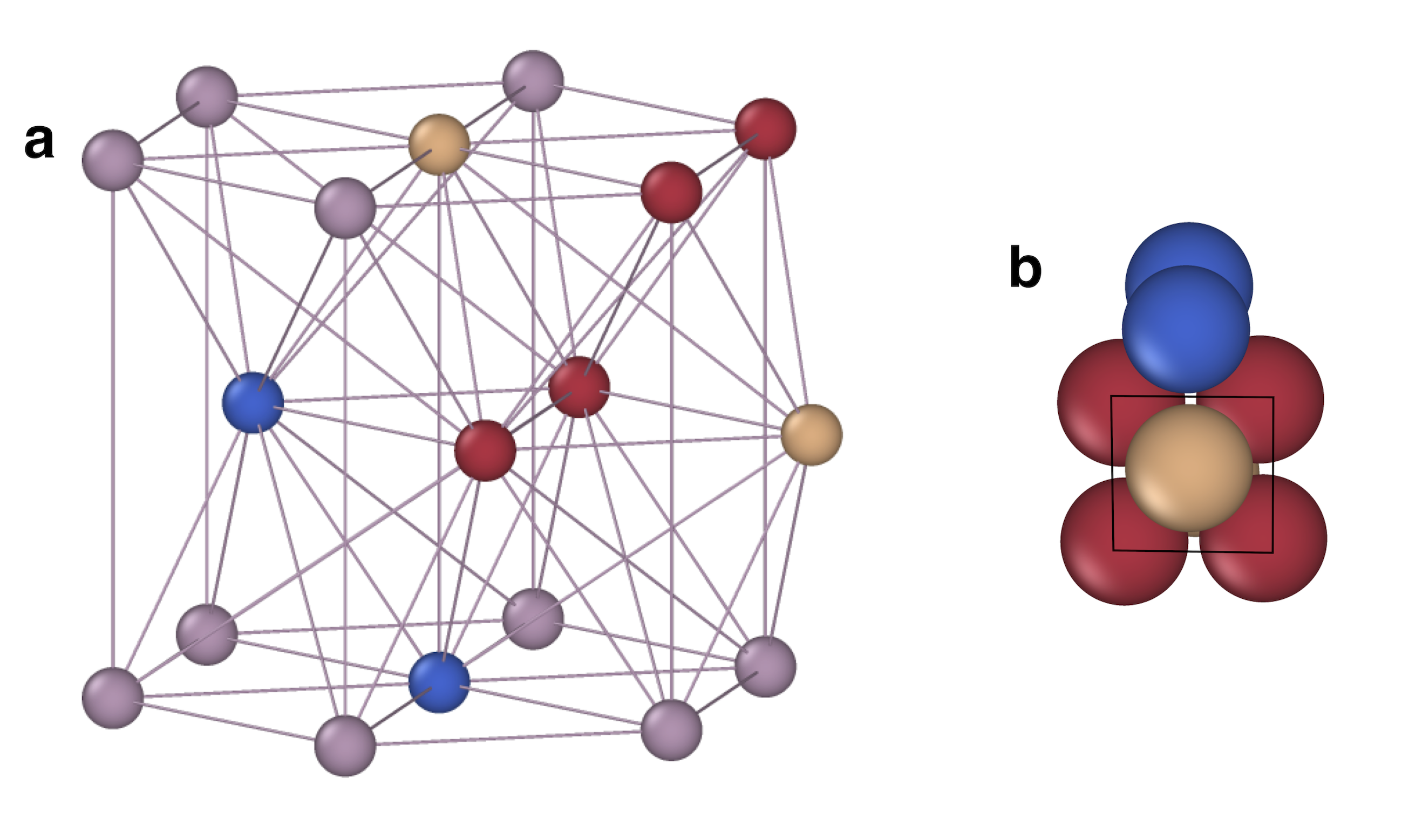}
\caption{\textbf{Geometrical relationship between an SD cluster and the hcp unit cell} (a) Unit cell of the hcp crystal with an additional particle belonging to the adjacent unit cell. An  defective SD cluster is identified in the unit cell of an hcp phase and the particles belonging to this defective SD cluster are coloured following the colour coding explained in the text. This SD cluster is termed defective as  one of the shifted particles (in blue)  is displaced with respect to an actual SD cluster. (b) The defective SD cluster resulting from the pattern in (a). From this viewpoint, it can be seen that one of the \emph{shifted} (blue) particles is displaced, as a result of the ``a-b-a-b" stacking of  hcp, differently from what is observed in the fcc case, where the stacking of the hexagonal planes follows the ``a-b-c-a-b-c" pattern.}
\label{fig:si_fig5}
\end{figure*}

This transformation is  sketched in Fig. S5. In Fig. S5a, we show the unit cell of an hcp crystal, with an additional particle on the right belonging to an imaginary adjacent unit cell. By including the latter particle, it is possible to find a pattern that resembles the SD cluster described in the main text.  We  therefore colour the particles with the usual colour coding, \emph{i.e.} \emph{ring} particles are coloured red, \emph{spindle} particles are coloured gold, and \emph{shifted} particles are coloured blue. As one of the shifted particles of this cluster is displaced with respect to an actual SD cluster, we denote this cluster as a defective SD.  Note that particles have been reduced in size so that the whole unit cell is visible. 

In Fig. S5b, we isolate only the particles belonging to this defective SD cluster and picture them with the actual colloid size. From the view point shown in this figure -- and comparing it with Fig. 3c and 3d of the main text -- it is evident that one of the \emph{shifted} (blue) particles is displaced with respect to its position in the SD cluster. Hence, the transition from an SD cluster to a defective SD cluster, which resembles a subunit of hcp, involves an additional displacement of one of the \emph{shifted} particles.  We therefore conclude that this additional displacement makes the propensity of fluid-like particles to crystallize into hcp lower than that of fcc.

